\documentclass[11pt,twoside]{article}   
\usepackage{latexsym} 
\usepackage{mathptm}
\usepackage{amsmath}
\usepackage{amssymb}
\usepackage{graphicx}

\def\beqn{\begin{eqnarray}}
\def\eeqn{\end{eqnarray}}

\begin{document}
 
\title{Comments on and Comments on Comments on Verlinde's paper ``On the Origin of Gravity and the Laws of Newton''}
\author{Sabine Hossenfelder \thanks{hossi@nordita.org}\\
{\footnotesize{\sl NORDITA, Roslagstullsbacken 23, 106 91 Stockholm, Sweden}}}

\date{}
\maketitle
\vspace*{-1cm}
\begin{abstract}
We offer some, hopefully clarifying, comments on Verlinde's recent claim that
gravity is an entropic force. A suitable identification of
quantities shows that both formulations of Newtonian gravity, the classical and the
thermodynamical one, are actually equivalent. It turns out that some additional assumptions made 
by Verlinde are 
unnecessary. However, when it comes to General Relativity there remain some gaps in the argument. 
We comment on whether this identification can be done also for electrostatics.
Finally, some thoughts on the use of this reinterpretation are offered.
\end{abstract}

\section{Introduction}

The purpose of this brief note is to offer some clarifying words on the logic of 
Verlinde's recent paper \cite{Verlinde:2010hp}. In his paper Verlinde showed
that (time-independent) Newtonian gravity is an entropic force and
claimed it to follow from a thermodynamical description that has certain holographic properties. 
Below, we will make this statement more precise. We will further show that 
the reverse is also true, i.e. it follows from Newtonian gravity that it can be described as
an entropic force with holographic properties. We are thus lead to conclude both
descriptions are equivalent and, after a suitable identification of quantities, 
the thermodynamical character of gravity is a reinterpretation.

Let us first state precisely what we mean. Let there be given a charge distribution $\rho$ which
is a density. Then, static Newtonian gravity can be characterized as follows:

\begin{quote} {\bf A:} There is a scalar field $\phi$ which obeys the Poisson equation $\nabla^2 \phi = 4 \pi G \rho$. 
A test-mass $m$ in the background field of a mass $M$ with field $\phi_M$ experiences a 
force $\vec F = m \vec \nabla \phi_M$.  \end{quote}
Here, $G$ is some coupling constant. Verlinde's proposal instead can be cast as follows
\begin{quote} {\bf B:} There are two scalar quantities $S$ and $T$ and a continuous
set of non-intersecting surfaces ${\cal S}$, the `holographic screens,' whose union
covers all of space ${\mathbb R}^3 = \cup {\cal S}$. The theory is defined by 
$2 G \int_{({\cal S})} \rho dV = \int_{\cal S} T dA ~\forall {\cal S}$, and the force acting
on a particle with test-mass $m$ is given by $F \delta x = \int_{{\cal S}} T \delta dS,$ where the
integral is taken over a screen that does not include the test-mass. \end{quote}
Here,  $({\cal S})$ denotes the volume with surface ${\cal S}$. The volume 
integral $\int_{({\cal S})} dV \rho$ is of course just the total mass $M$ inside that volume, and
the quantities $S$ and $T$ are interpreted as the entropy respectively temperature on the holographic 
screens. The $\delta x$ is a virtual variation on the location on the particle which
induces a change in the entropy on the screen, the details will become clear later.

The 
statement {\bf B} is not actually exactly what Verlinde claimed in \cite{Verlinde:2010hp}, in
particular he did not use the function $S$ itself but merely its differential $dS$. Verlinde further
defined another quantity $s$, a surface entropy density, with help of which $\delta dS = \delta s$,
and another quantity $N=GA$, the ``number of bits on the screen,'' with help of which $2 M = \int_{\cal S} T dN,$
the ``equipartition theorem.'' We will not use the quantities because we don't actually need them,
and our argument becomes leaner without them. The other
difference is that Verlinde defines the screens as being equipotential surfaces for $\phi$. 
However, in the thermodynamical approach $\phi$ should not enter the formulation of
the theory since it's what one wants to get rid of. Thus the above formulation ${\bf B}$ 
avoids referring
to the screens as equipotential surfaces. We'll come back to this later.

Before we get to the derivation, let's have a brief look at electrostatics, since
it is apparent that {\bf A} could equally have been about electric charges. To
avoid having to constantly distinguish two cases, we 
convert the charge to mass 
dimension one, and the potential to mass dimension zero by multiplying with powers of
the Planck mass $m_{\rm Pl}$ and subsume the remainder
in the coupling constant. For the case of gravity 
there's nothing to do. For
electrostatics we have $Q = m_{\rm Pl} \tilde Q$, $\phi = \tilde \phi/m_{\rm Pl}$, where
the quantities with tildes are the usual ones. This is just a rescaling, also used in \cite{Wang:2010px}, that will make
the following apply for both cases. We will thus generally refer to a charge and label it $q$, but
this charge could be a mass. We will on some places comment on the 
differences between electrostatics and gravity. Note that with the sign convention
of ${\bf A}$, which we use to be in agreement with \cite{Verlinde:2010hp}, a graviational 
potential $\phi$ is actually negative.  

We'll use the unit convention $\hbar = c= 1$ such that
$m_{\rm Pl}=1/l_{\rm Pl}$. We will consider the case with 3 spatial dimensions, such that 
$G=l_{\rm Pl}^2,$ though the number of dimensions doesn't matter much\footnote{The case with
only one dimension is pathological. To begin with because a point doesn't have a surface.}.

\section{It follows from Newtonian Gravity that it's an Entropic Force}

Now let us come to the derivation. Verlinde in his paper claimed to have shown ${\bf B} \Rightarrow {\bf A}$.
We will thus look at ${\bf A} \Rightarrow {\bf B}$, and start with the scalar field $\phi$ obeying 
the Poisson equation.

The field will have equipotential surfaces of codimension 1 where $\phi(\vec x)=$ constant. 
We identify these surfaces with the holographic screens. That's the first ingredient to {\bf B}. 
The values of these surface areas should
be a smooth function, in particular it won't have gaps. Each of these surface has a normal
vector in every point that we'll denote as $\vec n$. This normal vector might be ill-defined
in some points. For example a Lagrange-point has an indefinite normal vector. (If a particle was placed
exactly at this point, it would stay there.) 
Note that in the general case these surfaces will not be simply connected, but consist of
several pieces.

Let us assign the corresponding surface $A(\phi)$ to every value of $\phi$ on ${\cal S}$ and normalize it to a unit area $A_0=G$. We then define the following scalar function
\beqn
\quad S(\vec x) := -\phi(\vec x) \frac{A(\phi)}{2 G} + S_0 \quad,
\eeqn 
Here,
$S_0$ is some additive constant. 
Suggestively named $S$, in the case of gravity the function reduces to the black 
hole entropy for $r = 2MG$. We further
define another scalar quantity
\beqn
\quad T(\vec x) :=  \frac{1}{2\pi} \nabla_n \phi  \quad,
\eeqn 
which is the derivative in direction of the normal vector on the equipotential surface. For the case
of gravity again it reduces to the black hole temperature at the horizon. The quantity $T$ can
always be constructed where the normal vector is well-defined. For gravity where like charges
attract and there's only positive charges $T$ is positive definite, but for electric fields this
will not generally be the case.

It then follows by use of Gauss' law from the Poisson equation that on any equipotential surface
\beqn
\int_{({\cal S})} \rho dV = \frac{1}{4\pi G} \int_{\cal S} \nabla_n \phi dA  = \frac{1}{2 G} \int_{\cal S} T dA \quad, 
\eeqn
which is the second ingredient to ${\bf B}$. Note that this is not true for general surfaces since
the normal used to define the temperature will in general not be the normal of the surface.
Let us further recall the potential energy of the system is
\beqn
U = - \int \rho \phi dV  \quad,
\eeqn
which might contain divergent self-energy contributions that have to be suitably omitted (recall that with our
sign convention $\phi<0$). 
Now consider a test-particle with charge $q$ at location $r_0$ in the background of a charge $Q\gg q$ contained
within a compact volume. The background
field has potential $\phi_Q$ and the test-particle has the potential $\phi_q$ which by use of
the Poisson-equation is just the usual $\phi_q = - Gq/|r-r_0|$.
Now we ask what force we have to apply to change the test-particle's location by $\delta x$. It is
\beqn
\vec F \delta \vec x = \delta U = - \int dV \phi_Q \delta \rho = - \frac{1}{4 \pi G}\int dV \phi_Q \nabla^2 \delta \phi_q \quad, \label{from}
\eeqn
where the integral is taken over some volume that it includes the
test-particle, but not the background charges, i.e. it is outside some surface dividing both. 
Note that this has nothing to do with thermodynamics or holography whatsoever. 

We will
now repeat the steps used in \cite{Verlinde:2010hp}. First, we set the volume the integral is
taken over to be the volume inside an equipotential surface of $\phi_Q$ such that the test-particle
is outside that surface. We denote that surface ${\cal S}$, the volume inside $({\cal S})$ and
the volume outside ${\mathbb R}^3 \backslash ({\cal S})$. We start with rewriting the volume integral
in an integral over all of space minus the integral over the inside, and then transform the
volume over the inside by making use of Green's second identity (see Appendix). One obtains
\beqn
- 4 \pi G \delta U = \int_{{\mathbb R}^3}  \phi_Q \nabla^2 \delta \phi_q dV  - 
\int_{({\cal S})} \delta \phi_q \nabla^2 \phi_Q dV  + 
\int_{{\cal S}} \left(  \delta \phi_q \nabla \phi_Q  - \phi_Q  \nabla \delta \phi_q \right) dA~. 
\eeqn

Since we are integrating over an equipotential surface for $\phi_Q$, in the second term in the
surface integral we can pull out $\phi_Q$. Then we rewrite that surface-integral again in a volume
integral over $({\cal S})$ by making use of Gauss's law. We then see that it 
vanishes because $\phi_q$ does not have sources inside
the volume. The second term with the volume integral over the inside we rewrite again 
in a volume integral over ${\mathbb R}^3$ minus an integral over ${\mathbb R}^3 \backslash ({\cal S})$.
The integral over the outside vanishes because $\phi_Q$ does not have sources there. We are then
left with
\beqn
- 4 \pi G \delta U = \int_{{\mathbb R}^3} \left( \phi_Q \nabla^2 \delta \phi_q  - \delta \phi_q \nabla^2 \phi_Q \right) dV +
\int_{{\cal S}}  \delta \phi_q \nabla \phi_Q dA  ~.
\eeqn

For any finite volume, the remaining volume integral can be written as a surface integral making
use of Green's 2nd identity again. One
then sees that the integral vanishes if the surface is moved to infinity as long as
the sources are compact. Thus, we finally arrive at
\beqn
\vec F \delta \vec x = - \frac{1}{4 \pi G}\int_{\cal S} \delta \phi_q \nabla \phi_Q dA \quad, \label{thus}
\eeqn
where the surface integral could have been taken over any equipotential surface for the background
field $\phi_Q$ if the test-particle is not inside the volume enclosed by the surface. 
At such a fixed surface we have $2G \delta S = - A \delta \phi_q$ and the surface element is\footnote{This is the $\delta s$ in Verlinde's paper.}
$2G \delta dS = - \delta \phi_q dA$  thus
\beqn
\vec F \delta \vec x = \int_{\cal S} T  \delta d S ~. \label{one}
\eeqn
This completes the derivation ${\bf A} \Rightarrow {\bf B}$. 

One should note two points
about this derivation. One is that we have not made a variation of the surface. We have
instead made a variation of the potential on a the fixed surface, induced by wiggling the
location of the test-particle. The other point is that all steps from Eq. (\ref{from}) to
Eq. (\ref{thus}) can be reversed. For the final step from Eq.(\ref{thus}) to Eq.(\ref{one}), 
one uses the identification of the thermodynamical quantities with the Newtonian potential.

Since
$\delta \phi_q$ is not constant on ${\cal S}$ it is somewhat cumbersome but 
one can work out the integral on the 
right side of Eq.(\ref{thus}) for the case of two point-masses and obtain the usual Newtonian
gravitational force or Coulomb law respectively. This should not surprise us, since all we have 
done is using integral identities for harmonic functions. 

It is apparent that this derivation does not make use of the character of the charge. However,
let us look at a very simple case to see why the thermodynamic analogy does not make much
sense for electrodynamics. Consider two point-charges $q_1$ and $q_2$ very far away from each other,
and an equipotential surface with value $\phi_0$. If the charges are far enough away from each other, to good
precision the area of the equipotential surface is 
\beqn
A_{q_1} + A_{q_2} = 4 \pi \frac{G^2}{\phi_0^2} \left( q_1^2 + q_2^2 \right) ~.
\eeqn
Now we put the two charges very close together, and find that the area of the equipotential
surface changes to
\beqn
A_{q_1+ q_2} = 4 \pi \frac{G^2}{\phi_0^2} \left( q_1 + q_2 \right)^2 ~.
\eeqn
Now for gravity, two positive charges (masses) attract. In this process when
the masses approach each other one thus has an increase in area of the surface, proportional
to $q_1 q_2$.
Identifying the change of area for a constant value of $\phi$ with the
change in entropy, the entropy appropriately increases. However, for the
electromagnetic interaction it's opposite charges that attract, resulting in
a decrease of area or `entropy' since $q_1 q_2$ is now negative. 
This is in agreement with our previous observation
that for electrodynamics the `temperature' could be negative. 

Thus, while the derivation above can also be made for electrostatics, the interpretation
of $S$ and $T$ in thermodynamical terms does not seem meaningful. Only if the force is attractive 
between like charges the association with
an entropic force makes sense. If one starts with assuming that gravity
is an entropic force then, as pointed out by Smolin in \cite{Smolin:2010kk},
this means gravity is attractive between like charges. It does not exclude
however that there are negative gravitational charges, just that if there
were opposite charges one would have to do work to bring them closer together
since $dA,dS<0$ in that process.

\section{Newtonian Gravity follows from an Entropic Force Law}

For completeness, the following is a repetition of Verlinde's argument with some further remarks.
As mentioned previously, absent a gravitational potential we should not define the `holographic
screens' as its equipotential surfaces. One could think of instead defining them as
surfaces of constant entropy. However, from our previous identification of quantities
$2 G S = - \phi A(\phi)$, we see that a surface of constant $\phi$ is a surface of constant $S$, but
the reverse might not necessarily be true since a change in $\phi$ could be balanced by
a change in area. 

For a monopole (or far distance field), $\phi$ drops slower than the area of its equipotential
surfaces increases. The entropy thus increases monotonically with distance and 
thus the surfaces of constant $S$ do identify the surfaces of 
constant $\phi$. However, when we think of higher multipole moments, this might not
generally be true\footnote{At least it is not obvious to me. Comments are welcome.}. But we have additional
information in $T$. With it, we can define a screen ${\cal S}$ as a surface with constant
entropy that fulfils the relation $2G \int_{({\cal S})} \rho dV = \int_{\cal S} T dA$ in every point. 
On these screens with area $A({\cal S})$, we can then get back the Newtonian potential by
defining it as $\phi = - 2G S /A({\cal S})$. Consequently, if we make a small change to
$S$ on a constant surface, we have $\delta \phi = - 2G \delta S/A({\cal S})$. 

Having all quantities necessary for ${\bf A}$, the rest of the argument is then an exact reversal of the steps in the previous
section, with one additional subtlety. If one knows that $4 \pi G \int_{(\Sigma)} dV \rho = \int_\Sigma \nabla_n \phi dA$
holds for every surface $\Sigma$ with normal vector $\vec n$, then one can easily conclude that $\rho$ must fulfill the Poisson
equation. However, we only have this relation for equipotential surfaces. The missing information is in the
shape of the surfaces themselves. This can be seen as follows. The charge distribution $\rho$ and the normal on 
the (known) surface constitutes a
case of a Neumann boundary condition and if the field fulfills the Poisson equation it is uniquely specified up to
a constant. 
If there was another, different, solution not fulfilling the equation, it would have to differ from the solution
to the Poisson equation in at least one point. But this would mean the equipotential surfaces had to be shifted
and thus cannot be. 

Taken together, we have thus shown that ${\bf A} \Leftrightarrow {\bf B}$.

\section{Remarks}

The number $N$ of `bits' on the screen made no appearance here. In particular, 
making the ``identification for
the temperature and the information density on the holographic screens'' \cite{Verlinde:2010hp} might be ``natural'' but 
is unnecessary. In any case if one uses
that $A=G N$ then to obtain Eq.(\ref{one}), we kept $A$ and thus $N$ constant. Since the missing variation
with respect to $N$ was criticized in \cite{Gao:2010yy,Culetu:2010ua}, let us point out there is nothing
inconsistent about this, one just has to be careful about stating which quantities are held fix (as usual
in thermodynamics). Since we did not need $N$, the equipartition theorem also does not play a
role. 

Further, for the above it was unnecessary to use that $\Delta S$ is proportional to $\Delta x$. This is
just a shortcut to arrive at Newton's gravitational force law, but not necessary for the general case
which deals with the variation of the entropy on a screen induced by varying the position of the
test-particle. Since the relation $\Delta S \propto \Delta x$ does not enter the derivation, it is moot to 
dissect its meaning, discuss the constant of proportionality, or the appropriateness in general cases. One
also does not actually need any knowledge about the Unruh-effect, though this knowledge helps to give
meaning to the definition of the temperature.

It should be emphasized that the direction ${\bf B} \Rightarrow {\bf A}$ is the more complicated
direction of the equivalence. This is partly due to the identification of the potential. But mostly
it is due to ${\bf A}$ being a description available in the whole volume and all its surfaces,
while ${\bf B}$ makes use only of specific surfaces. That both are equivalent nevertheless is
a consequence of the specific properties of solutions to the Poisson equation.

 In the later sections of \cite{Verlinde:2010hp} the argument is 
extended from Newtonian gravity to General Relativity (GR) in a spacetime with a time-like killing vector. 
From the argument presented there, one sees that ${\bf A} \Rightarrow {\bf B}$ can be extended
to GR. However, the assumptions for the reverse claim, that 
gravity follows from thermodynamics, rely heavily on
knowledge about GR already: Instead of defining the gravitational
potential from thermodynamical quantities, the temperature is defined making use of
the GR generalization of the gravitational potential. It enters
the (previously defined) killing vector. The definition of
screens still has the same problem as in the Newtonian case. Later it is moreover 
used that the total mass inside a volume can be rewritten
in terms of the stress-energy-tensor, and in addition that this stress-energy-tensor 
is covariantly conserved. One also has to use the equivalence principle when going 
into a locally Minkowski frame. Finally, one might wonder if, absent something like a 
space-time, it was appropriate to use derivatives to begin with, or if these should
not also `emerge' in some sense. 
  
\section{Discussion}

In any case, it presently does not seem impossible to fix the gaps pointed out in the previous
section, so let us for now consider there
was indeed a thermodynamical description equivalent to GR. 
This then raises two questions: What does this mean? and What is it good for?

In the previous sections we avoided an interpretation of the reformulation suggested
by Verlinde. It remains to be clarified whether there is more justification for referring
to the quantities in thermodynamical terms.
Of course one wonders what it's supposed to mean that any point in spacetime is now
associated with a temperature. For the case of a black hole horizon there is
an actual particle emission associated to the temperature. Should we now expect 
arbitrary screens to also emit particles?

Well, first a screen that is not a black hole horizon does not have a
problem emitting particles. It is clear where they come from: they come from
the next screen closer to a source and can be traced back to the source which
is a sink of field lines. The
problem with a horizon is just that all paths crossing 
it are classically forbidden. However, the temperature of any such screen is
ridiculously small and entirely negligible for all practical purposes. Consider a proton with a
mass of $m_{\rm p} \approx$~GeV. Assuming it has a  has a typical size of $\approx 1/m_{\rm p}$,
it has a surface temperature of $T \approx m_{\rm p} (m_{\rm p}/m_{\rm Pl})^2\approx 10^{-25}$eV.
Asking how long it would take for the proton to noticeably lose a quantum of energy,
we can as usual integrate $dM/dt \approx R^2 T^4$. One finds that it would take 
about $10^{100}$ times the current age of the universe for the proton to decay in that 
fashion. We can safely forget about that\footnote{That this number is so large compared
to the black hole temperature is because $T$ scales with the third power of the
mass rather than with the inverse of the mass as is the case for a black hole horizon.}.

In any case, if observable consequences of the reinterpretation are such remote, 
what is it good for? Clearly, the equivalence of two formulations for the gravitational 
interaction is not the interesting aspect. As long as both descriptions are the same claiming
``the redshift must be seen as a consequence of the entropy
gradient and not the other way around'' \cite{Verlinde:2010hp} is merely words. The interesting aspect of the
reformulation would be to make use of it in regimes where the equivalence 
might no longer hold. The thermodynamical description might provide a bridge
to a statistical mechanics description of a possibly underlying theory.

 We know that gravity is not easy to quantize, and there are cases we
cannot describe by use of GR. The question is whether in these
cases classical gravity might cease to be meaningful, but the thermodynamic
relations continue to hold, thus making them accessible to us. However, before one
starts thinking about quantum gravity one should maybe try to first reproduce
at least quantum mechanics in a classical background field. The next step could
then be asking for the description of the gravitational force exerted by a superposition
state, which is already a case where a classical description of the particle's
gravitational field fails.

Another remark, mentioned before in \cite{Gao:2010yy}, is that we do have a fundamental theory for
electrodynamics that is extremely well tested. Since formally the identification with
thermodynamics can also be made for electrostatics it might be illuminating to
find out whether it continues to hold under quantization. And then it should be
clarified what to do about solutions of the Laplace solutions, i.e. how to reproduce
gravitational waves. 

Finally, let's come back to the claim that according to this reformulation, gravity has a
holographic character. Since the thermodynamical interpretation is
just an equivalent rewriting there's no more holography in it then there was in
Newtonian gravity or electrostatics till last year. Arguably one could say though that harmonic functions
do have `holographic' properties, encoded by Gauss's and Stoke's integral identities that
we heavily made use of.  Rather than speaking
of holography, it might be
more appropriate to refer to such a theory that does have bulk degrees of freedom but allows
for a holographic description as `holographic friendly,' meaning it might possibly be
extendable to regimes where bulk degrees of freedom are indeed not available\footnote{Lee Smolin, private communication.}. 

\section*{Acknowledgements}
I thank Robert Helling, Stefan Scherer, Lee Smolin and Erik Verlinde for comments and helpful discussions.

\section*{Appendix}

For convenience, here is Gauss's law
\beqn
\int_{({\Sigma})} \nabla^2 \psi dV = \int_{\Sigma} \nabla_n \psi dA \label{Gauss}~.
\eeqn
and Green's 2nd identity
\beqn
\int_{({\Sigma})} \left( \psi \nabla^2 \phi - \phi \nabla^2 \psi\right) dV = \int_\Sigma \left( \psi \nabla_n \phi - \phi \nabla_n \psi \right) dA ~. 
\eeqn
where, as previously, $(\Sigma)$ denotes the volume inside a surface $\Sigma$, and
$\nabla_n$ is the derivative in direction of the surface's normal vector. $\psi$ and $\phi$ are both scalar fields.

\end{document}